\documentclass{article}

\newcommand{\authname}[1]{{\footnotesize\sffamily\bfseries #1}}
\newcommand{\authadd}[1]{{\small\rmfamily\itshape #1}}

\usepackage{amsfonts}
\usepackage{amsmath}
\usepackage{amssymb}
\usepackage{graphicx}
\usepackage{subfig}
\usepackage{theorem}
\usepackage{xspace}
\usepackage{bbold}
\usepackage{color}
\usepackage{float}
\usepackage{epsf}
\usepackage{natbib}
\usepackage{setspace}

\newcommand{\set}[1]{\ensuremath{\{#1\}}}
\newcommand{\formula}[1] {\begin{displaymath} \begin{array}{l} #1 \end{array} \end{displaymath} \xspace}
\newcommand{\markedformula}[2] {\begin{equation} \label{#1} \begin{array}{l} #2 \end{array} \end{equation} \xspace}
\newcommand{\Cic} {\mathcal{C}}

\newcommand{\Spec}[1]{\mathcal{#1} \xspace}

\theoremstyle{break}

\newtheorem{example}{Example}

\theoremstyle{plain}

\newtheorem{definition}{Definition}

\newtheorem{statement}{Statement}

\nocite{*}
\nocite{*}
\begin{document}

\sloppy

\title{The Stabilizing Role of Global Alliances in the Dynamics of Coalition Forming}

\author{\authname{Galina Vinogradova}\footnote{galina.vino@gmail.com}\\[2pt]
\authadd{CREA - Center of Research in Applied Epistemology, Ecole
Polytechnique}\\
\authadd{Palaiseau, France}\\
\and
\authname{Serge Galam}\footnote{serge.galam@cnrs-bellevue.fr}\\[2pt]
\authadd{CNRS - National Center of Scientic Research} \\
\authadd{Paris, France} }

\maketitle

\begin{abstract}

Coalition forming is investigated among countries, which are coupled
with short range interactions, under the influence of external
fields produced by the existence of global alliances. The model
rests on the natural model of coalition forming inspired from
Statistical Physics, where instabilities are a consequence of
decentralized maximization of the individual benefits of actors
within their long horizon of rationality as the ability to envision
a way through intermediate loosing states, to a better
configuration. The effects of those external incentives on the
interactions between countries and the eventual stabilization of
coalitions are studied. The results shed a new light on the
understanding of the complex phenomena of stabilization and
fragmentation in the coalition dynamics and on the possibility to
design stable coalitions. In addition to the formal implementation
of the model, the phenomena is illustrated through some historical
cases of conflicts in Western Europe.


\end{abstract}

\section{Introduction}

This work is devoted to the study of stabilization in coalition
forming in a collective of individual actors under the influence of
external fields. The model rests on the natural model of coalition
forming \cite{GAVNM} inspired from the Statistical Physics' model of
Spin Glasses \cite{SGM}, through which the system of countries is
compared to a collection of interacting spins -- tiny magnetic
dipoles that interact with each other and align themselves in a way
to attain the most "comfortable" position, the one that minimizes
their energies. While the presentation addresses the coalition
forming in an aggregate of countries, the discussion and the results
can be applied to any type of political, social or economical
collectives where the association of actors takes place based on
their bilateral propensities.

This work subscribes to the growing field of modeling complex social
situations using Statistical Physics \cite{CFLSF} which has started
over thirty years ago with \cite{SYY}. Later, a study of collective
decision making combining Social Psychology hypotheses with recent
concept of Statistical Physics \cite{MG} set the frame of using spin
Hamiltonian. Then, the coalition as a form of aggregation among a
set of actors (countries, groups, individuals) has been studied
using concepts from the theory of Spin Glasses \cite{Axel, FVS, GC,
SDO, Flo, APIM}. Various social applications of the model were
suggested \cite{SPC, TBISCF, GAVNM}. The dynamical analogue of this
model was introduced in \cite{GAVDP}.

The model of coalition forming among countries leans on the
existence of strong and static bilateral geographic-ethnic
propensities linking the countries. Those propensities have emerged
during the ongoing historical interactions between neighbor
countries and appear to favor either cooperation or conflict. Their
spontaneous and independent evolution have produced an intricate
circuit of bilateral bonds which causes contradictory tendencies in
the simultaneous individual searches for optimal coalitions. Due to
stronger interactions with a common ally, conflicting countries may
be brought to cooperate momentarily despite their natural tendency
to conflict. Such a situation produces an endeavor of the concerned
countries to escape from the unfavorable cooperation leading to
instabilities, which in turn produce a break down of the current
coalitions inducing the formation of new ones.

The origin of such instability is twofold, either coming from
spontaneous fluctuations or directed by external attraction towards
a global alliance. The extremely disordered dynamics of coalitions
and fragmentation in Western Europe in past centuries belong to the
first kind, while the building up of the Soviet and Nato global
alliances is of the second kind.

In this work we aim to study the instability of coalition forming
among countries, which, in contrast to physical entities, are
rational actors that are able to maximize their individual benefits
through a series of choices within a decentralized maximization
process.

On this basis, coalitions are formed through the short range
interactions between the countries -- the attraction or repulsion
based on the unalterable historical bonds between them. According to
the principle that " the enemy of an enemy is a friend", the
countries are assumed to ally to one of two competing coalitions.

Allying to the same coalition is unfavorable to the countries which
went through historical rejection. As a result, such countries seek
to affiliate with the opposite coalitions. Alternatively, allying to
the opposite coalition is unfavorable to friendly countries.
Countries which belong to the same coalition are expected to
cooperate even if their natural propensity is to conflict. Such a
contradiction results in a potential instability.

Our previous study \cite{GAVNM} focused on studying the effects of
instabilities arising in the coalition forming among rational actors
as a function of the bonds structure, the optimal and non-optimal
stabilizations as well as the robustness of the stability.

In this work the model is extended to investigate the mechanisms by
which the setting of a global alliance produces attraction in an
aggregate of individual countries otherwise connected by their
natural bilateral propensities. In particular, the focus is on how
those new interactions can eventually stabilize the intrinsically
unstable process of coalition forming keeping the short range nature
of interactions. Global attraction is ensued from a global external
field set over the system of countries, which in turn polarizes the
countries' interests and produces incentive unifications under two
opposing \emph{global alliances}. The resulting coalitions are
affected by the net bilateral balance between the new motivations
and the traditional historical ones.

We focus on how the interactions produced by attraction to global
alliances overwhelm the current instability among the countries. The
results provide new theoretical tools that enable to measure the
efficiency of a global attraction in forming stable alliances, as
well as to theoretically design new effective global attractions
that can yield stability.

The study of stabilization of coalitions using global alliances was
started in \cite{SPC}. The authors describe spontaneous formation of
economic coalitions given a random distribution of propensity bonds,
and illustrate new exchanges between the countries incited by the
global alliances. Those exchanges, along with an additional
parameter of economical and military pressure, are viewed as the
ones that produce additional bilateral propensities yielding new
stable coalitions.

In the current work, we develops further the research on coalition
forming under a global external field. We address the stabilization
by unique factors -- such as economical, political, social,
ecological, as well as by multi-factor stabilization, where the
influence of several independent factors is equiprobable. Based on
the new formulation, we investigate the remarkable historical cases
of conflicts in Western Europe.

The multi-factor stabilization is an innovative concept both in
Political Sciences where it explains the complexity of coalition
forming, and in Statistical Physics where it illustrates how a
stable disorder arises from an anti-ferromagnetic coupling achieved
by the interlocking of two opposite ferromagnetic states. Some forms
of such mixed phases of ferromagnetism have been studied in
\cite{CSGM}.

\section{Background -- The Natural Model and Instability}\label{sec_model}

The Spin Glass model in Statistical Physics is an idealized model of
bulk magnetism represented by a collection of interacting spins --
atoms acting as a tiny dipole magnet with a mixture of ferromagnetic
and anti-ferromagnetic couplings. Those magnets interact with each
other seeking to align themselves parallel or anti-parallel in order
to minimize their energies. The collection of spins forms a
disordered material in which the competing interactions cause high
magnetic frustration -- changes of spins at no energy cost, with a
highly degenerate ground state.

\begin{figure}[!ht] \centering
\includegraphics[width=2.3in]{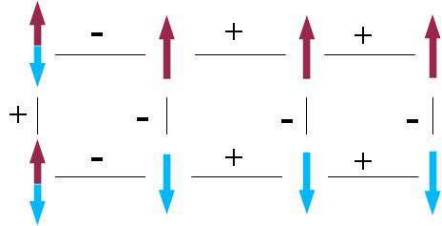} \caption
{Ising model of $8$-spins with mixed pair interactions. The pair
propensity bonds are denoted by $+$ or $-$, and states of the spins
are denoted by the arrows. Frustrated spins are marked by both up
and down arrows. This Spin Glass phase yields an unstable
disorder.}\label{spin_glass}
\end{figure}

The Ising model of a random bond magnetic system can be described as
follows. The model consists of $N$ discrete variables
$\set{S_i}_{1}^{N}$, called spins, that can be in one of two states
\emph{up} or \emph{down}. Figure (\ref{spin_glass}) shows
schematically the case of $8$ spins with identical amplitude of the
propensity bonds located on a lattice and interacting at most with
their nearest neighbors. The spins for which a shift of the state
cost no energy are defined as frustrated.

The natural model of coalition forming is formally identical to the
Ising model with pure or mixed anti-ferromagnetic couplings in a
particular geometry of the lattice. the model considers a system of
$N$ countries whose historical interactions have defined propensity
bonds between them, which are either positive (ferromagnetic-like)
or negative (anti-ferromagnetic-like). To each country labeled with
an index $i$ ranging from $1$ to $N$, is attached a discrete
variables $S_i$ which can assume one of two state values $S_i=+1$ or
$S_i= -1$. The values correspond to the country's choice between the
two possible coalitions. The same choice allies two countries to the
same coalition, while different choices separate them into the
opposite coalitions.

Combination of states of all the countries $S= \set{S_1, S_2, S_3,
\dots, S_N}$ forms a state configuration that defines an allocation
of coalitions. Here, by symmetry, both configuration $S$ and it's
inverse $-S = \set{-S_1, -S_2, -S_3, \dots, -S_N}$ define the same
coalitions.

Bilateral propensities $J_{i,j}$ have emerged from the respective
mutual historical experience between the countries $i$ and $j$. The
propensities measure the amplitudes and the directions of the
exchange between two countries -- cooperation or conflict.
$J_{i,j}$, which is symmetric, is zero when there are no direct
exchanges between the countries.

Product $J_{ij} S_iS_j$ measures the benefit or the gain from the
interactions between both the countries as a function of their
choices. Aimed to maximize this measure, the countries seek to ally
to the same coalition when $J_{ij}$ is positive and to the opposing
ones, otherwise. Thus, depending on the direction of the primary
propensity, the conflict can be beneficial to the same extent as the
cooperation.

The sum of the benefits from all the interactions of country $i$ in
the system makes up the net gain of the country:
\markedformula{countr_gain}{H_i(S) = S_i\sum_{j\ne i} \hspace{0.1in}
J_{ij}S_j.} Thus, a configuration $S$ that maximizes the gain
function defines the country's most beneficial coalition setting.

\begin{figure}[!ht] \centering
\includegraphics[width=1.5in]{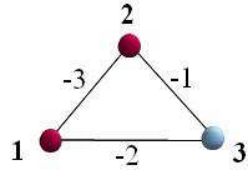} \caption
{Triangle of three conflicting countries $1$, $2$, $3$ with negative
mutual bonds. }\label{trian_natural_m}
\end{figure}

For the sake of visualization, we depict the system of countries
through a connected weighted graph with the countries in the nodes
and the bilateral propensities as the weights of the respective
edges (see Figure (\ref{trian_natural_m})). We take red (dark) color
for the $+1$ choice and blue (light) color for the $-1$ choice.

The total gain of the system of countries is identical to the
Hamiltonian of an Ising random bond magnetic system which represents
the energy of the system. For a configuration $S$, we have for
system's gain : \markedformula{syst_gain}{\mathcal{H(S)} =
\frac{1}{2} \sum_i \hspace{0.1in}  {H_i(S)}.}

In physical systems, the Hamiltonian -- the function that determines
the physical properties of the spin system, is precisely concerned
with minimization of the system's energy. This physical analogy
allows to address the bilateral propensities between the countries
as mean of maximization of the countries' individual gain
(minimization of their energy) and as the principal guide in the
coalition forming.

A major difference between the model of spins and the model of
rational countries is the long horizon rationality of the countries
in contrast to the spins, which are only able to foresee only the
immediate effect of their shifts. Countries have the ability to
maximize their individual benefits through a series of planned
changes while assuming possible losses in the intermediate phase.

The Ising model, indeed, can be represented through the natural
model where the countries' rationality is limited to observation of
an immediate gain, optimizing only their local maximums.

When the most beneficial coalition configurations of different
countries do not coincide, the maximization of individual gains
induces competitions for the beneficial associations. Among the
countries with complete rationality which are aware of attainability
of a better configuration, those competing interactions cause
endless instability in the system. However, the system may remain
stable when some actors have limited rationality -- being unaware of
possibility to attain a better configuration, they are satisfied
having reached a local maximum.

\begin{figure}[!ht]
\centering
\includegraphics[width=1.5in]{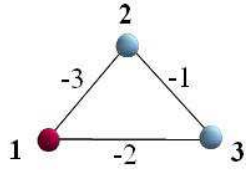} \caption
{The triangle of conflict. The triangle is stable when the countries
$2$'s and $3$'s rationality is limited to immediate improvements;
any change cause a loss in their gain. The triangle is unstable when
the countries are fully rational: their most beneficial
configurations do not coincide and the countries, being able to
forecast an improvement, make changes that impair the gain in the
immediate steps. }\label{trian_nat_config}
\end{figure}

Figure (\ref{trian_nat_config}) shows the triangle of conflict in a
configuration where it is stable when the countries $2$'s and $3$'s
rationality is limited to immediate improvements; any change cause a
loss in their gain. The triangle is unstable when the countries are
fully rational: their most beneficial configurations $S_1=
\set{\set{+1, -1, -1}, \set{-1, +1, +1}}$, $S_2= \set{\set{-1, +1,
-1}, \set{+1, -1, +1}}$, $S_3= \set{\set{-1, -1, +1}, \set{+1, +1,
-1}}$, do not coincide. Here, at any coalition configuration, at
least two of the countries improves their gain when the other
changes. As a result, aiming in their best configurations and being
able to forecast an improvement at any step, the countries may make
changes that impair the gain in the immediate steps.

It is interesting to note that for the case of equal propensities
over the edges, the triangle of conflict is unstable for any limited
rationality actors, including the spins, due to zero-value gain
produced in the cyclic geometry resulting in no-cost frustrations.

\begin{definition}[Instability of the System of Countries]
The system of countries is said to be unstable if in any
configuration of the countries' states there is a country which is
able to forecast an improvement of its gain.
\end{definition}

Negative product on a circle means an unpaired negative coupling
where two neighbors are found to be connected both though positive
and negative branch in the circle. This creates an everlasting
competition between the neighbors for the exclusive arrangement to
ally with the positive branch. The countries thereby continuously
shift their respective choices producing the instability.

In Statistical physics the necessary condition of instabilities in
Spin Glasses \cite{GT} reads that \markedformula{stab_spins}{$
\emph{the instability implies the existance of a closed circle
of spins connected with the bonds on which}$\\
$\emph{the product of total bonds is negative}.$} Indeed, the Spin
Glasses' instability is a result of frustrations, and a negative
circle can appear to be stable as soon as a shift increases the spin
energy preventing the spin flop.

In contrast to the Spin Glass model where changes are limited to the
spontaneous no-cost fluctuations, in the natural model where the
instability is due to the rationality of actors, changes may impair
the immediate gain. In the theoretical interpretation of the model
where the complete rationality of all countries is assumed, the
terms (\ref{stab_spins}) are also the sufficient condition of the
instability in the model -- the condition of endless competitions
among the countries for the beneficial configurations.

Formally, the theoretical terms of instability in the natural model
are as follows. Denote a circle of countries by $\Cic$ and the
countries composing the circle by $1,2, \ldots, k$.
\markedformula{stab_countr}{$\emph{If there is a closed circle of
countries on which the product of total propensities is negative},$\\
\hspace{2in} \Pi_{i,j \in \Cic} \hspace{0.1in} p_{ij} < 0 \hspace{0.02in}\\
$\emph{then the system is unstable}.$}

Let us remark that, in the theoretical interpretation of the model
where the complete rationality of all countries is assumed, the
instability is not value-dependent but is determined by the signs of
the propensities -- the distribution is such that involves a
negative circle. At the same time, any local maximum strictly
depends on the propensities' values.

\section{Global Alliance Model Of Coalition Forming}

Global alliance model starts from a global principle which
represents an external field polarizing the interests of the
countries. This leads to the emergence of two opposing global
alliances. The countries attach themselves to one or to the other
based on their pragmatic interests with respect to the global
principle. The new interactions, while favor either cooperation or
conflict, stimulates contributions to the countries' mutual
propensities. The new prospectives unify or separate the countries
based on the pragmatic motivations which in combination with the
historical concerns allow other distributions of coalitions.

Here we address the role of the global alliances in forming of
stable coalitions among the countries or other rational actors.
Whether the system is unstable or there is an optimal or local
maximum stable configuration, the new exchanges between the
countries incited by the global alliances impact the stability. For
the sake of simplicity of presentation, we assume the extensive
rationality of the countries. While in such theoretical
interpretation the instability is not value-dependent, the effect
from the globally generated additional propensities on the stability
is subject to the values of primary propensities. Therefore, in
spite of the extensive rationality of the countries, in the
presentation we address the model with arbitrary range of values.

Let us define the global alliance model formally. The global
alliance unifies the countries that support the global principle,
while its opponents are unified under the opposing global alliance.
Denote the two alliances by $M$ and $C$. A country's individual
disposition to the alliances is determined by the countries'
cultural and historical experiences and is expressed through the
parameter of \emph{natural belonging}. The natural belonging
parameter of country $i$ is $\epsilon_i = +1$ if the country has
natural attraction to alliance $M$, $\epsilon_i = -1$ for $C$.

By making a choice among the two possible state values $S_i=+1$ and
$S_i= -1$, country $i$ chooses to belong to either alliance $M$ or
$C$. The choice of $+1$ allies the country to alliance $M$ and the
choice of $-1$ allies it to alliance $C$. Any particular
distribution of two countries among the alliances creates new
interactions between the countries whose directions depends on the
natural disposition of the countries. Namely, if the countries are
attracted to the opposing alliances, the exchange will be negative
as soon as they ally to the same alliance.

Those new exchanges between any two countries $i$ and $j$ define
additional propensity between the countries. The propensity is the
amplitude of the exchange $G_{ij}$ in the direction $\epsilon_i
\epsilon_j$ that favors either cooperation or conflict. For the
purpose of this presentation we assume that the exchange amplitudes
are unchanged.

The overall propensities between the countries, involving both the
historical inclinations and the propensities resulting from the new
exchanges, are determined as follows
\markedformula{add_prop}{p_{ij}= J_{ij} + \epsilon_i \epsilon_j
G_{ij}.} Respectively, the net gain of country $i$ is
\markedformula{glob_gain}{H_i = S_i\sum_{j \ne i} \hspace{0.1in}
{(J_{ij} + \epsilon_i \epsilon_j G_{ij})S_j}.}

Thus, in the presence of external incentives of the global
alliances, the couplings between the countries obtain new guidance.
The countries adjust their states to the best benefit with regards
to the new propensities. The new choice of coalition is determined
by both spontaneous reactions and planned interactions, which enable
coupling based on a planned profit.

\section{Stabilization Of The System By Additional Factors}

Here we address the stabilization of coalition forming in the system
where rational countries have no optimal configuration of
coalitions, and where, as result, the spontaneous stabilization can
not be attained. Global alliances based interactions in such systems
enable stable coalitions among the actors even if they remain of
short-range nature. Such interactions, however, being a complex
superposition of several factors of countries' objectives, must
satisfy particular stability constraints.

\subsection{The Uni-Factor Stabilization}

Consider two opposing global alliances $M$ and $C$ in a system of
$N$ countries. A particular factor of the countries' interests
produces specific dispositions to the global alliances which
encourage new exchanges between the countries. The appropriate
amplitudes of the exchanges enable the stabilization among the
countries, the \emph{uni-factor stabilization}.

With respect of unique factor of stabilization, the necessary and
sufficient condition stability (reformulated terms
~(\ref{stab_countr})) is that \markedformula{glob_total_prop_eq}
{$\emph{A system is stable if and only if for any circle $ \Cic $ in the system,} $ \\
\Pi_{i,j \in \Cic} \hspace{0.1in} (J_{ij} + \epsilon_i \epsilon_j
G_{ij}) \ge 0.}

Now we state the existence of a stable coalition within the global
alliance model.
\begin{statement}
The presence of global alliances, regardless of the global principle
that produced them, enables a stable coalition among countries.
\end{statement}

In order to prove this statement, let us first observe that the
product of the additional propensities $p^{\Spec{G}}_{ij}=
\epsilon_i \epsilon_j G_{ij}$ on any circle is always positive.
Indeed, given circle $\Cic$, \formula{\Pi_{i,j \in \Cic}
\hspace{0.1in} G_{ij} \hspace{0.1in} \epsilon_i \epsilon_j =
\Pi_{i,j \in \Cic} \hspace{0.1in} G_{ij} ( \epsilon_1 \epsilon_2
\epsilon_3 \ldots \epsilon_{k}) ^2  = \Pi_{i,j \in \Cic}
\hspace{0.1in} G_{ij}.} This implies that on any circle, the number
of negative couplings produced by the global alliances is even. If
the system is unstable, than there is at least one negative circle.
We define the new interaction amplitude as follows. For each couple
$i,j$ whose $\epsilon_i \epsilon_j < 0$ we take $G_{ij} = 0$ if the
primary propensity is negative and $G_{ij} = 2 |J_{ij}|$ for the
positive original coupling. When $\epsilon_i \epsilon_j
> 0$, we take $G_{ij} = 2 |J_{ij}|$.

Making the new propensities negative for the negative global
couplings and positive for the positive ones, guarantees that there
is an even number of negative couplings on the circle. This remains
invariant for each circle in the system, which implies that the
construction produces non-negative product on any circle in the
system. The stability condition (\ref{glob_total_prop_eq}) holds
true which concludes the proof of the statement.

\subsubsection{A Case of the England-Spain-France Triangle}

A typical examples of the uni-factor stabilization is stabilization
of the triangle of England, Spain and France (\ref{engspfr}) during
historical events of $1584$ \cite{JISP}.

\begin{example}[ Stabilization in the $ESF$ Triangle of Conflict by the Religious Factor ]\label{conf_tri_unistab}

Against the background of sequence of wars in the old Europe, the
countries attained stability when in $1584$ Catholic Spain and
France formed an alliance against Protestant forces, the most
notable of which were settled in England.

In order to illustrate historical example using the global alliance
model, we describe the propensities between the countries from
"negative" to "positive" through mixed ones. Attaching to them
numerical values with respect to their relative strength, taking
"neutral" as $0$.

Accounting for the historical relationship between England, Spain
and France, we take the propensities as "neutral-negative",
"negative" and "highly negative". There numerical interpretations,
as shown in Figure (\ref{engspfr}), are arbitrary values that aim to
account for a relative strength of the interactions.
\begin{figure}[!ht]
\centering
\includegraphics[width=1.5in]{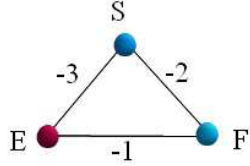} \caption
{Triangle of England ($E$), Spain ($S$) and France ($F$), the
$ESF$-conflicting triangle. }\label{engspfr}
\end{figure}

By $M$ and $C$ we denote the two opposing global alliances -- the
countries in $M$ choose unification into a "European union" and
those in $C$ are against the unification. With respect to the
religious factor, Catholic Spain and France were naturally
associated to $M$ ($\epsilon_S, \epsilon_F = 1$), while Protestant
England was associated to $C$ ($\epsilon_E = -1$). Then, $\epsilon_S
\epsilon_F = 1$, $\epsilon_E \epsilon_S = \epsilon_E \epsilon_F =
-1$, and the overall propensities between the three countries are:\\
$p_{SE}= -3 - G_{SE}$, $ p_{EF} = -1 - G_{EF}$ and $p_{SF}= -2 +
G_{SF}$.

Solving the inequality \markedformula{trian_stab_term}{(-3
-G_{SE})(-1 - G_{EF})(-2 + G_{SF}) \ge 0} yields the constraint the
new interaction amplitudes $G_{SE}, G_{EF}, G_{SF}$ must satisfy in
order to stabilize the triangle. Since $G_{EF}$, $G_{SE}$ and
$G_{SF} > 0$, the only root of the respective equality is $G_{SF} =
2$. The solution space is $G_{SF} \ge 2$, as depicted in Figure
(\ref{trian_sol}), represents a three-dimensional space of the
independent additional propensities.

\begin{figure}[!ht] \centering
\includegraphics[width=2in]{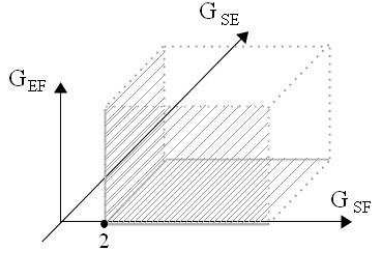} \caption
{Three-dimensional solution space of the independent additional
propensities in the uni-factor stabilization of the $ESF$- triangle
of conflict.}\label{trian_sol}
\end{figure}

In the historical example, coalition of Spain and France against
England implies that the amplitude of their new interaction belonged
to the solution space. The respective stable configuration is $S =
(+1, -1, -1)$, as shown in Figure (\ref{engspfr_stab})) where
$G_{EF}$, $G_{SE}$ are taken to be $0$ and $G_{SF}$ to be $3$, so
that the corresponding total propensities become $-1$,$-3$ and $1$.

\begin{figure}[!ht] \centering
\includegraphics[width=1.5in]{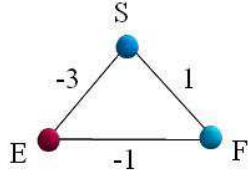} \caption
{The global alliances model of the $ESF$-triangle stabilized by the
religious factor in configuration $~{S= (+1, -1, -1)}$. Here,
$G_{EF} = G_{SE} =0$ , $G_{SF} = 3$, so that the respective
resulting total propensities are $-1$,$-3$ and $1$.
}\label{engspfr_stab}
\end{figure}

\end{example}

It is interesting to observe that:
\begin{statement}
Any system of countries in the global alliance model with a unique
factor of interests is reducible to a stable system represented in
the natural model.
\end{statement}
Indeed, given a system in the global alliance model, let us define
the new state variable to be $\tau_{i} = \epsilon_i S_i$. The
variable takes a value of $\set{ +1,-1}$. Then, the hamiltonian
$H_i$ of country $i$ can be written in the terms of the new state
variables as \formula{H_i = \sum_{i \ne j} \hspace{0.1in}(
J_{ij}S_iS_j + G_{ij}\epsilon_i \epsilon_jS_iS_j) = \sum_{i\ne j}
\hspace{0.1in} (J_{ij}\epsilon_i \epsilon_j + G_{ij})\tau_i \tau_j.}
Here, since $G_{ij}$ is positive, some choice of
$\set{G_{ij}}_{i,j}$ produces the propensities that guarantee a
stable system.

\subsection{The Multi-Factor Stabilization}

Taking into account only one factor of countries' interests would be
too restricting -- along with religious interests, the global
principle may impact economical, ecological, moral, political or any
other interest and concern. Distinct interests simultaneously
influence the interactions between the countries in different ways.
They modify the countries' propensities by aggregating the
corresponding independent interactions -- economical, political and
others.

Let us define formally the multi-factor form of the global alliance
model through two coexisting factors of interests, denoted by
$\Spec{G}$ and $\Spec{K}$ respectively. Within each factor, a
country has independent natural disposition to the global alliances.
Therefore, each country has two independent natural belonging
parameters associated with the factors. For country $i$, this is
$\epsilon_i = +1$ if within factor $\Spec{G}$ the country naturally
belongs to $M$. Similarly, $\beta_i = +1$ within factor $\Spec{K}$.
For the global alliance $C$, $\epsilon_i = -1$ and $\beta_i = -1$
respectively.

We denote by $G_{ij}$ the amplitude of the exchanges between the
countries $i$ and $j$ on factor $\Spec{G}$, and by $K_{ij}$ the
amplitude on $\Spec{K}$. Then, the total new propensity between the
countries $i$ and $j$ is the superposition of those directed
exchanges on the two factors: $p^{\Spec{G}}_{ij} = \epsilon_i
\epsilon_j G_{ij}$ and $p^{\Spec{K}}_{ij} = \beta_i \beta_j K_{ij}$.
The two-factor form of the global alliance model superposes the
spontaneous interactions of the natural model with the intended
interactions based on the two-dimensional choice among the global
alliances: $p_{ij}= J_{ij} + \epsilon_i \epsilon_j G_{ij} + \beta_i
\beta_j K_{ij}$. The net gain of country $i$ is \formula{H_i = S_i
\sum_{j\ne i} \hspace{0.1in}  S_j( J_{ij} + G_{ij}\epsilon_i
\epsilon_j + K_{ij} \beta_i \beta_j) .}

In order to illustrate the multi-factor stabilization, we turn again
to the Example (\ref{conf_tri_unistab}) of stabilization of the
$ESF$- conflicting triangle.

\subsubsection{ Multi-factor Stabilization of the England-Spain-France Triangle}\label{conf_tri_multistab}

We assume, in addition to the religious factor $\Spec{G}$ in the
$ESF$- conflicting triangle, that there is an economical factor
$\Spec{K}$. In this golden age Spain had a pronounced disinclination
to any economical unification with its old enemies, while England
and France remarked the advantages of such unification. Therefore,
the respective parameters of natural belonging on the economical
factor $\Spec{K}$ are $\beta_S = -1$ and $\beta_E, \beta_F = 1$.
With respect to the economical factor, the overall propensities
between the countries are : $p_{SE}= -3 - G_{SE} - K_{SE}$, $p_{EF}=
-1 - G_{EF} + K_{EF}$, and $ p_{SF}= -2 + G_{SF} - K_{SF}$.

\begin{figure}[!ht] \centering
\includegraphics[width=3in]{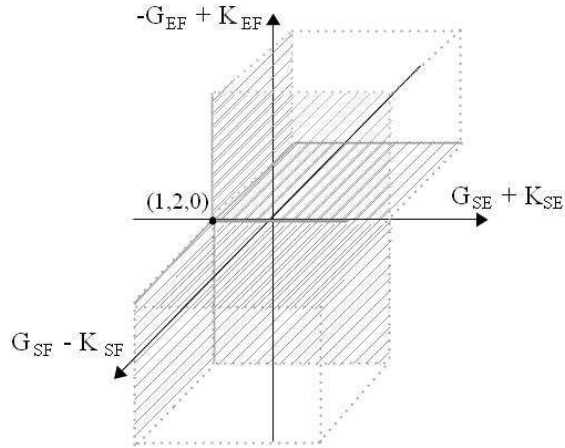} \caption
{three-dimensional solution spaces of the independent additional
propensities in the two-factor stabilization of the
$ESF$-triangle.}\label{trian_2_sol}
\end{figure}

Solutions of inequality \markedformula{trian_multi_fact_eq}{\Pi_{i,j
\in \Cic} \hspace{0.1in} p_{ij}= (-3 - G_{SE} - K_{SE})(-1 - G_{EF}
+ K_{EF})(-2 + G_{SF} - K_{SF}) \ge 0} yield the exchange amplitudes
that guarantee stability of the $ESF$-triangle in the multi-factor
form. Since $G_{SE} + K_{SE} \ge 0$, the solution must satisfy
$-G_{EF} + K_{EF} \ge 1$ and $G_{SF} - K_{SF} \le 2$, or $- G_{EF} +
K_{EF} \le 1$ and $G_{SF} - K_{SF} \ge 2$ (see Figure
(\ref{trian_2_sol})).

\begin{figure}[!ht] \centering
\includegraphics[width=1.5in]{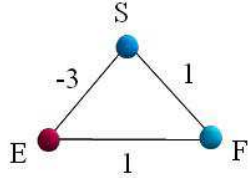} \caption
{The global alliances model of the $ESF$-triangle in the
multi-factor case. Here, $G_{SE} = 0$, $G_{EF} = 2$ and $G_{SF} =
3$, so that the respective total propensities become $1$, $-3$, $1$.
The system remained unstable because the global exchange amplitudes
did not satisfy the terms of stability -- the circle remained
negative. }\label{engspfr_multi}
\end{figure}

In the historical reality of this period of the countries, the
economical factor $\Spec{K}$ could not produce interactions as
strong and significant as the exchanges on the religious factor.
That is why $K_{EF} < 1 + G_{EF}$ and $K_{SF} < G_{SF} + 2$ which
have prevented the $ESF$-triangle to reach the stability until
religion took a secondary place conceding importance to economics.
See Figure (\ref{engspfr_multi})), where $G_{SE}$ is taken to be
$0$, $G_{EF}$ to be $2$ and $G_{SF}$ is $3$, so that the respective
total propensities become $1$, $-3$, $1$.

It worth to notice that in the multi-factor form, a system in the
global alliance model can be no more interpreted as a system in the
natural model as soon as the choice of at least two countries
differs on at least two factors. Still, the general multi-factor
case can be reduced to the two-factor form of the global alliance
model: one of the factors unifies the amplitudes of all the positive
coupling and the other unifies those of all the negatives ones.

Therefore, with no restriction on the generality, the multi-factor
form of global alliance model can be studied within the case of two
coexisting factors. This also explains the fact that in the majority
of cases, only two camps of opposing concerns play the crucial role
in the coalition forming.

\section{Physical Interpretation of the Multi-Factor Stabilization}

\begin{figure}[!ht] \centering
\includegraphics[width=2.3in]{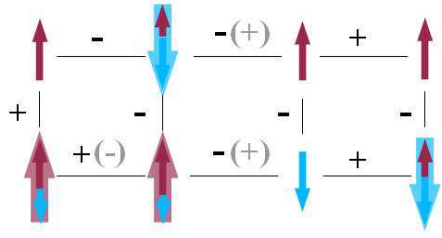} \caption
{Ising model of $8$-spins, initially mixed negative and positive
pair interactions (highlighted by grey color), is stabilized by
mixing of two ferromagnetic states. Each spin's absolute direction
(marked by the thick arrows) is the average of those ferromagnetic
directions. Among the two opposite directions, either one of them
dominates or the two eliminate each other, thus neutralizing the
ferromagnetic states on the spin. The Spin Glass phase yields a
stable disorder.}\label{spin_glass_two_fact}
\end{figure}

In the context of Statistical Physics, the multi-factor
stabilization is equivalent to the superposition of unstable
disorder of a spin glass with two stable orders (two factors) of
ferromagnetic states which split the spins in two directions (two
alliances). Each spin's absolute direction is the average of those
ferromagnetic directions, as shown in Figure
(\ref{spin_glass_two_fact})). Among the two opposite directions,
either one of them dominates or the two eliminate each other, thus
neutralizing the ferromagnetic states on the spin. In the figure,
thick arrows indicate the absolute directions of the spins, and thin
arrows show their ferromagnetic directions.

The multi-factor stabilization of coalition forming is an innovative
concept both in Political Sciences and in Statistical Physics. In
the former, it explains the multitude of elements influencing the
coalition forming. In the later, it shows how in a frustrated system
a stable disorder is achieved from interlocking of two ferromagnetic
states of opposite directions with anti-ferromagnetic coupling among
them.

\section{Multi-factor Stabilization in Western Europe}

Here we attempt to illustrate the formation of Italian state within
the context of the global alliance model. It is known that, given a
system from the reality, it is hard to obtain exact numerical values
of the propensities in the system. Once such values are known we can
explain the transitions and predict resulting configurations with
arguable precision. Having no such values, we still can provide some
analysis based on estimated values of the propensities extracted
from the historical chronicles. Running the model with those values
allows to analyze and explain the transitions and the result
configuration. This can not be done based only on the canonical
representations of historical events.

Let us illustrate the Italian unification in $1856$ - $1858$, where
four countries were involved: Italy, France, Russia and Austria
\cite{EPR} and \cite{UIT}. The period from the end of 18th till the
middle of 19th century was marked by the series of European wars
including the French invasion of Italy where Austrian and Sardinian
forces had to face French army in the War of the First Coalition,
The War of the Fifth Coalition of Austria against French Empire.

In 1852, the new president of the Council of Ministers in an Italian
region Piedmont, Camillo di Cavour, had expansionist ambitions one
of which was to displace the Austrians from the Italian peninsula.
An attempt to acquire British and French favor was however
unsuccessful.

Then, Napoleon III, who had belonged to an Italian family
originally, decided to make a significant gesture for Italy. In the
summer of 1858, Cavour and Napoleon III agreed to cooperate on war
against Austria. According to the agreement, Piedmont would be
rewarded with the Austrian territories in Italy (Lombardy and
Venice), as well as the Duchies of Parma and Modena, while France
would gain Piedmont's transalpine territories of Savoy and Nice.

Despite the Russian help in crushing the Hungarian Revolt in 1849,
Austria failed to support Russia in the Crimean War of the middle of
1850s. Therefore, Austria could not count on Russian help in Italy
and Germany. Alexander II has agreed to support France in a fight
with Austria for the liberation of the Italians, though only by
showing up the army on its borders with Austria. It appeared to be
enough to force the Austrians withdrew behind the borders of Venice.

However, the conquest of Venice required a long and bloody mission,
which may cause revolts and threaten Napoleon III's position in
France. In the private meeting with Franz Joseph, together they
agreed on the principles of a settlement to the conflict according
to which the Austrians have to cede Savoy and Nice to the French,
yet would retain Venice. The Russian was indignant at this turn of
France.

\begin{figure}[!ht] \centering
\includegraphics[width=2.3in]{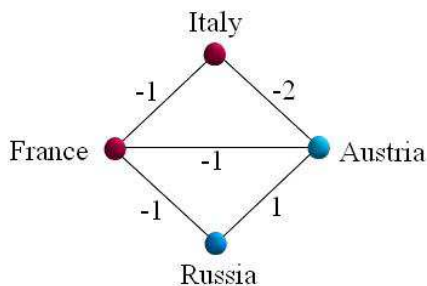} \caption
{The unstable system of France, Russia, Italy, Austria with their
relative primary propensities, 1856-1858.}\label{fiar}
\end{figure}

Let us reproduce the historical chronicle presented above with the
help of our model. The initial states of the countries with their
primary propensities are shown in Figure (\ref{fiar}).

The value of propensities indicating the relative strength of
primary interactions between the countries are taken from "negative"
to "positive" through mixed ones with the respective numerical
values taking "neutral" as $0$ are shown in Figure \ref{fiar}. Thus,
the historical relationships between the two absolutist monarchies
Russia and Austria are estimated as "neutral" with $J_{RA} = 1$.
Italy and Russia having no noticeable political relationship are
"neutral" to each other. The Franco-Russian relationship built up
during the French Revolutionary and the Napoleonic Wars is are
rather "neutral-negative" with $J_{FR} = -1$, as well as the
interactions of France with Italy and Austria who had experienced
series of military conflicts. The opposition between Italy and
Austria tied to the mutual territorial claims is estimated to be
"significantly-negative" with $J_{IA} = -2$).

Figure (\ref{fiar}) shows the system of the countries in its natural
model. The model has two negative circles and so is unstable which
appears through the historical changes before the rise of the
Italian question. The instability originates from the fact that
France gets identical benefits from the alliance with Russia and
Italy as from the opposing alliance with Austria.

An external field in the model results from the principle of
independent state of Italy. The respective opposing global alliances
are $M$ which associates the countries that support the independence
of Italy, and $C$ which unifies the countries opposing the
independence.

Here, two respective factors influencing the historical series of
events must be distinguished: external politics with the military
goals, and internal politics involving the social concerns of the
countries (their governing classes). Denote the two factors by
$\Spec{G}$ and $\Spec{K}$ respectively.

With respect to their external goals, Italy and France, as well as
Russia, agree to the relevance of an independent state of Italy.
Yet, in the social concerns the governing classes of France, Russia
and Austria agree in their rejection of socialist ideas springing
over all the Italy. Therefore, the respective parameters of the
countries' natural belonging to the alliances are distributed as
follows. With the natural belonging parameter $\epsilon$ referring
the external goals and $\beta$ referring the internal social
politics, for France $\epsilon_F = +1 $ and $\beta_F= -1$, for Italy
$\epsilon_I = +1$ and $\beta_I= +1$, for Russia $\epsilon_R = +1 $
and $\beta_R= -1$, and for Austria $\epsilon_A = -1 $ and $\beta_A=
-1$.

The global alliance motivated propensities are given is the
following chart:
\begin{center}
    \begin{tabular}{| l | l | l | l | l | l |}
    \hline
    Propensity & F-I & I-A & F-R & F-A & R-A\\ \hline
    Primary & -1 & -2 & -1 & -1 & 1 \\ \hline

    On $\Spec{G}$ & $G_{FI}$ & $- G_{IA}$ & $G_{FR}$ & $ -G_{FA} $ & $- G_{RA}$ \\ \hline

    On $\Spec{K}$ & $- K_{FI}$ & $- K_{IA}$ & $ K_{FR}$ & $ K_{FA} $ & $ K_{RA}$  \\ \hline

    \hline \noalign{\smallskip}
    \end{tabular}
\end{center}

The historical chronicle of the four countries is concluded in three
phases: a phase of no global alliances, or the natural model phase,
and two phases of global alliances rose due to the Italian question,
where in the first one the external and military concerns come to
picture and in the second one the internal social concerns rise over
the countries.

As we have seen in Figure (\ref{fiar}), the system in its natural
model is unstable, where France fluctuates between Russia and
Austria.

Let us evaluated the amplitudes of the military exchanges between
the countries through numerical values providing the relative
magnitudes of interactions. Russia has equally "moderate" interest
in military cooperation with both France and Austria, with $G_{FR} =
2$ and $G_{RA} = 2$. Italy and France are "strongly" interested in
the conflict having Italian land at stake, $G_{FI} = 4$ and $G_{IA}
= 4$, while the interest between Austria and France is
"moderately-strong" with $G_{FA} = 3$. A sympathy of Russian to
Italian state comes up in the "basic" interest, $G_{RI} = 1$. The
new propensities between the countries with respect to the external
politics interests are shown in the following table:
\begin{center}
    \begin{tabular}{| l | l | l | l | l | l | l |}
    \hline
    Propensity    & F-I & I-A & F-R & F-A & R-A & R-I\\ \hline
    Primary       & -1  & -2  & -1  & -1  & 1   &  0  \\ \hline

    On $\Spec{G}$ & 4   & -4  & 2  & -3  & -2 & 1 \\ \hline

    Total         & 3  & -6  & 1  & -4  & -1 & 1 \\ \hline
    \hline
    \end{tabular}
\end{center}

\begin{figure}[!ht] \centering
\includegraphics[width=2.3in]{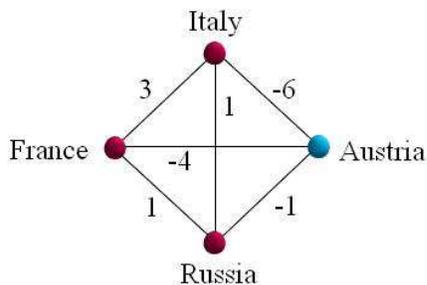} \caption
{France, Russia, Italy and Austria, 1856-1858, with the new military
propensities. It forms a stable system with the coalition of France,
Italy and Russia against Austria.}\label{fiar1}
\end{figure}

As result of the interactions the system obtain a new shape shown in
Figure (\ref{fiar1}). Here, absence of negative circles allow a
perfectly stable coalition of France, Italy and Russia against
Austria .

However, the social aspect of the internal politics of the countries
dramatically interferes with the stability. The relative amplitudes
of the consequent exchanges can be estimated as follows. Due to the
political insularity of Russia where serfdom still prevailed over
large part of the country, the amplitudes of all its exchanges on
the social aspect are "negligible", $K_{FR} = 0$ and $K_{RA} = 0$.
France and Austria had a "strong" involvement in the subject, with
$K_{FA} = K_{FI} = K_{IA} = 4$. The new propensities between the
countries are shown in the table.
\begin{center}
    \begin{tabular}{| l | l | l | l | l | l | l |}
    \hline
    Propensity    & F-I & I-A & F-R & F-A & R-A & R-I \\ \hline

    Primary       & -1  & -2  & -1  & -1  & 1   &  0  \\ \hline

    On $\Spec{G}$ & 4   & -4  & 2  & -3  & -2 & 1 \\ \hline

    On $\Spec{K}$ & -4  & -4  & 0   & 4   & 0    & 0 \\ \hline

    Total         & -1 & -10  & 1  & 0 & -1 & 1 \\
    \hline
    \end{tabular}
\end{center}

\begin{figure}[!ht] \centering
\includegraphics[width=2.3in]{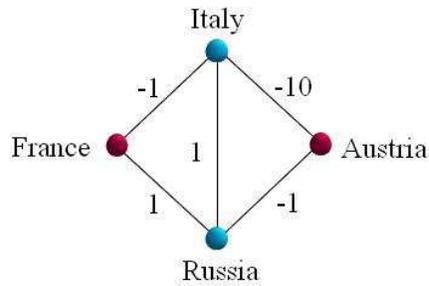} \caption
{France, Russia, Italy, Austria, 1856-1858, with the new
propensities on both the military and social factors. The result of
the war for liberation of Italy is the instability in a new shape. }
\label{fiar2}
\end{figure}

The result system with the French change in favor of cooperation
with Austria is shown in Figure (\ref{fiar2}). As we can see the
modified system includes three negative circles. The change of
France put Russia in an unfavorable position moving it away from a
most beneficial coalition configuration. At the same time, Italy and
Austria found themselves in a satisfactory state.

\section{Conclusions}

Coalitions in a collective of individual rational actors such as
countries, when formed spontaneously are rare to stabilize. The
probability that the system becomes stable vanishes exponentially
with the size of the system. In reality, stabilization among
countries as rational actors is more likely to happen under the
external incentive of global alliances and is more practical. The
impact of the global principle on the economical, political, social
or any other factor of the countries' interests produces new,
intended, interactions between the countries. In contrast to the
spontaneous primary interactions, those interactions are intended in
the sense that they are based on the directed view of the
countries's needs and interests. Superposed with the spontaneous
ones, the interactions guarantee the stabilization once their
amplitudes satisfy the constraints of positive circuits of
propensities.

One of the interesting directions for further research in the
context of the global alliance model is to study the general effect
from the global attractions, that is the general interaction
amplitudes. While some global attractions represent efficient
mediators, the others may be less successful or even harmful with
respect to the system's stability. Because either they become
obsolete, or provide no sufficient motivations, or they acts with
harmful intentions, the global alliance may fail to stabilize an
unstable system, and even may destabilize a stable system. It is
interesting to study those effects from general perspective of the
system's total gain (energy), which can be reduced or augmented by
the global alliances. The study should help shed the new light on
conflicts in post colonial Africa and  Middle East which, being
under the influence of external fields, continuously cycle in series
of contentions.


\begin{thebibliography}{}



\bibitem[Vinogradova \& Galam (2012)]{GAVNM} Vinogradova G. \& Galam S. (2012).
Rational Instability in the Natural Coalition Forming, \emph{Physica
A: Statistical Mechanics and its Applications}, 392 (2013)
6025--6040.


\bibitem[Binder \& Young(1986)]{SGM}Binder, K. \& Young, A.P. (1986). Spin-Glasses: experimental facts, theoretical concepts, and open
questions, \emph{Review of Modern Physics}, 58, 801--911

\bibitem[Castellano \& Fortunato\&  Loreto (2009)]{CFLSF}
C. Castellano \& S. Fortunato \& V. Loreto (2009). Statistical
physics of social dynamic, \emph{Reviews of Modern Physics}, (81)
591--646, 2009.


\bibitem[Galam \& Gefen \& Shapir(1982)]{SYY} Galam S. \& Gefen Y. \& Shapir Y. (1982).
Sociophysics: A new approach of sociological collective behavior,
\emph{British Journal Political Sciences}, 9, 1--13.

\bibitem[Galam \& Moscovici(1991)]{MG}Galam, S. \& Moscovici, S. (1991). Towards A Theory Of Collective Phenomena: Concensus And Attitude
Changes In Groups, \emph{European Journal Of Social Psychology}, 21,
49--74


\bibitem[Axelrod \& Bennett (1993)]{Axel} Axelrod, R. \& Bennett, D.S. (1993). A landscape theory of
aggregation, \emph{British Journal Political Sciences}, 23, 211--233


\bibitem[Galam (1996)]{FVS}  Galam S. (1996).
Fragmentation Versus Stability In Bimodal Coalitions,
\emph{Physica}, A, 230, 174--188

\bibitem[Galam (1998)]{GC} Galam, S. (1998). Comment on A landscape theory of aggregation,
\emph{British Journal Political Sciences}, 28, 411--412


\bibitem[Matthews (2000)]{SDO} Matthews R. (2000).
A Spin Glass model of decisions in organizations, \emph{Business
Research Yearbook, G. Biberman, A. Alkhafaji (eds), Saline,
Michigan: McNaughton and Gunn}, 7, 6

\bibitem[Florian \& Galam (2000)]{Flo} Florian, R. \& Galam, S. (2000). Optimizing
conflicts in the formation of strategic alliances, \emph{Eur. Phys.
J. B }, 16, 189--194

\bibitem[Tim Hatamian (2005)]{APIM} Tim Hatamian G. (2005).
On alliance prediction by energy minimization, neutrality and
separation of players, \emph{arxiv.org/pdf/physics/0507017}

\bibitem[Galam (2002)]{SPC} Galam S. (2002). Spontaneous Coalition Forming. Why Some Are Stable?,
\emph{Springer-Verlag Berlin Heidelberg 2002, S. Bandini, B.
Chopard, and M. Tomassini (Eds.):ACRI 2002, LNCS 2493}, 1--9

\bibitem[Gerardo \& Samaniego-Steta \& del Castillo-Mussot \& Vazquez (2007)]{TBISCF}
Gerardo G. N. \& Samaniego-Steta F. \& del Castillo-Mussot M. \&
G.J. Vazquez (2007). Three-body interactions in sociophysics and
their role in coalition forming,
 \emph{Physica}, A, 379, 226--234

\bibitem[Vinogradova (2012)]{GAVDP} Vinogradova G. (2012).
Correction of Dynamical Network's Viability by Decentralization by
Price, \emph{Journal of Complex Systems}, 20, 1, 37--55

\bibitem[Van Hemmen(1982)]{CSGM} Van Hemmen J.L. (1982).
Classical Spin-Glass Model, \emph{Physical Review Letters}, 49, 6


\bibitem[Toulouse (1977)]{GT}Toulouse, G.
  (1977). \emph{Theory of the frustration effect in
Spin Glasses: I}, Comm. on Physics, 2

\bibitem[Israel (1997)]{JISP} Israel J. I. (1997).
Conflicts of Empires: Spain, the Low Countries, and the Struggle for
World Supremacy, 1585--1713.


\bibitem[Hales (1954)]{EPR} Hales E.E.Y. (1954).
A Study in European Politics and Religion in the Nineteenth Century,
\emph{P.J. Kenedy}.

\bibitem[Beales \& Biagini (2003)]{UIT} Beales D. \& Biagini E. (2003).
The Risorgimento and the Unification of Italy, \emph{Longman}, 2nd
ed.



\end{thebibliography}
\end{document}